\documentclass[12pt]{article}
\usepackage{latexsym} \usepackage{epsf}
\usepackage{epsfig}
\usepackage{a4}
\usepackage{amsfonts}
\usepackage{showlabels}
\usepackage[colorlinks]{hyperref}
\usepackage{color}
\newcommand{\ba}{\begin{array}}
\newcommand{\ea}{\end{array}}
\newcommand{\be}{\begin{equation}}
\newcommand{\ee}{\end{equation}}
\newcommand{\nn}{\nonumber}
\newcommand{\bea}{\begin{eqnarray}}
\newcommand{\ena}{\end{eqnarray}}
\newcommand{\beas}{\begin{eqnarray*}}
\newcommand{\enas}{\end{eqnarray*}}

\newcommand{\mt}{\mathcal}

\begin{document}

\begin{center}
{\Large{ Explicit
R-matrices for inhomogeneous 3D chiral Potts models: Integrability and the action
formulation for IM}}
\end{center}

\begin{center}{\bf
Sh. Khachatryan,  {\footnote{e-mail:{\sl shah@mail.yerphi.am}} A.
Sedrakyan {\footnote{e-mail:{\sl sedrak@mail.yerphi.am}}}
 }}
 \\ A.~Alikhanyan National Science Laboratory (YerPhI), 
 \\ Alikhanian Br. str. 2, Yerevan 36, Armenia
\end{center}

\begin{abstract}
We construct the exact spectral parameter dependent vertex R-matrix for
the classical 3D $\mathcal{N}$-state chiral Potts models, convenient for considering the model
 in context of the Bethe ansatz. The R-matrix is defined on the $\mathcal{N}^4$
dimensional space $V_\mathcal{N}\otimes V_\mathcal{N}\otimes  V_\mathcal{N}\otimes V_\mathcal{N}$, appropriate for consideration by means of the cube-equations defined in
\cite{sfks}. We
present the 2D quantum spin Hamiltonians for general case and, at $\mathcal{N}=2$, a fermionic lattice
action representation corresponding to 3D Ising's statistical model.
\end{abstract}

{\small\tableofcontents}
\newpage

\section{Introduction}

The Ising models are the ones of the  most investigated statistical integrable models in one and two
dimensions \cite{Ons, Kauf,Baxt,ShML,KW,IZ,JW}, and until now there are intensive investigations for
elucidating the problem for three and more dimensions \cite{Pol, Sedrakyan-1984,KavS,FSS,Beta,Igbal-2020,IMc}. A direct
integrable generalization of the 2D IM is the $\mathcal{N}$-state chiral Potts model \cite{FZ,ZFG,YMP,YPB,YPBS,chBS}. In this work we are demonstrating the 3D version of the
 vertex $R$-matrices for general inhomogeneous $\mathcal{N}$-state Potts models, starting from the
$\mathcal{N}=2$ case of IM. The explicit integrablity conditions are not  investigated yet, however as all the 2D projection matrices are the solutions to the Yang-Baxter equations, hence the investigating of the models on the surfaces would bring us to deal with (1+1)d integrable models.

The second section devotes to the investigation of the  3D  Ising model by means of the
 technique which we have used in \cite{SA} for the investigation of 2D spin models.
As it was done in \cite{SA},
we define here an explicit form of the $R$-matrix, starting from the classical
statistical weights of the model.  Evaluating appropriate unitary local transformations of the states and operators we establish the operator form of $R$.  Although the states are defined on the
vertices and the interaction is considered as around a cube, however in this form the model can be
considered as a "vertex" model.  This is the direct analog of the 2D situation in \cite{SA}, where we  used vertex-like Yang-Baxter equations for the $R$-matrices defined on the faces of the lattices (with the spin-states situated on the vertexes). In the Bethe ansatz concept  two neighboring R-matrices defined such way  in the
transfer matrices have common vertices, but no common links (for 2D cases) or faces (for 3D cases).  One can try to employ here the cube equations presented in \cite{sfks}, which are appropriately defined for the vertex kind four-state R-matrices.

 In the next section we generalize our approach to $\mathcal{N}$-state chiral Potts model. As it is known, for 2D case, the chiral Potts model is the integral generalisation of the 2DIM \cite{FZ}-\cite{GSG}.  The algebraic structure of the corresponding 3D four-state R-matrix is presented. The corresponding 2D quantum spin Hamiltonian operators are presented also.

 Then, in  fourth section, we are formulating a
scalar fermionic action for 3DIM model ($\mathcal{N}=2$), representing the
partition function in the coherent-state fermionic basis  as a
continual integral. Free fermionic conditions are presented. The
fermionic interpretation of  IM is not new \cite{Kauf,FSS,KW,Baxt,Pol,KavS}, for 3D case
see for example
the super-symmetric non-interacting strings model in \cite{Pol},
or in \cite{KavS}. As it was stated in \cite{Pol}, the model is
reduced to the two-dimensional supersymmetric Liouville theory,
so, at critical point  the 3D Ising model should be described by a
conformal field theory. The information of statistical
characteristics are obtained  by Monte Carlo simulations, and
there are numerous works using the conformal field conception
\cite{Beta,IMc}. Approximate value of the homogeneous coupling
constant at the critical point has been found here in the free fermionic
limit and the difference from known approximate value  of the
constant presented in the mentioned works  is $\approx 0.05$.

 \section{The 3D weight matrix and the corresponding R: IM}
 \setcounter{equation}{0}

 Here we investigate the one of the simplest 3D spin models - 3DIM. The statistical weight of the 3D Ising model, defined on the cell of the cubic lattice $N\times N\times N$, can be written as follows
 \bea
 W_{\bar{\sigma}_{\alpha_1}\bar{\sigma}_{\alpha_2}\bar{\sigma}_{\alpha_3}\bar{\sigma}_{
 \alpha_4}}^{\bar{\sigma}_{\beta_1}\bar{\sigma}_{\beta_2}\bar{\sigma}_{\beta_3}\bar{\sigma}_{\beta_4}}=
 e^{J_1(\bar{\sigma}_{\alpha_1}\bar{\sigma}_{\alpha_2}+\bar{\sigma}_{\alpha_3}\bar{\sigma}_{\alpha_4}+
 \bar{\sigma}_{\beta_1}\bar{\sigma}_{\beta_2}+\bar{\sigma}_{\beta_3}\bar{\sigma}_{\beta_4})+J_2(\bar{\sigma}_{\alpha_1}\bar{\sigma}_{\alpha_3}
 +\bar{\sigma}_{\alpha_2}\bar{\sigma}_{\alpha_4}+
 \bar{\sigma}_{\beta_1}\bar{\sigma}_{\beta_3}+\bar{\sigma}_{\beta_2}\bar{\sigma}_{\beta4})+
 J_3\sum_i^4\bar{\sigma}_{\alpha_i}\bar{\sigma}_{\beta_i}},\nn\\\label{wi}
 \ena
 where $\bar{\sigma}_i=\pm 1$ are the projections of the spin operator on the site indexed by $i$. Then the statistical sum reads
   \bea Z=\prod_{i_x,i_y,i_z=1}^{N} W_{\;\;\;{\bar{\sigma}_{2\vec{a}_i}\;\;\;\bar{\sigma}_{2\vec{a}_i+a_x}\;\;\;\bar{\sigma}_2{\vec{a}_i+a_y}\;\;\;
 \bar{\sigma}_{2\vec{a}_i+a_x+a_y}}}^{\bar{\sigma}_{2\vec{a}_i+a_z}\;\bar{\sigma}_{2\vec{a}_i+
 a_x+a_z}
 \;\bar{\sigma}_{2\vec{a}_i
 +a_y+a_z}\;\bar{\sigma}_{2\vec{a}_i+\vec{a}}}.\label{Zi}
  \ena
  Here the projections of the vector $\vec{a}=\{a_x,\;a_y,\;a_z\}$ are the spacings of the 3D cubic lattice in the corresponding spacial directions, the sites on the lattice are denoted by $a_i=\{i_x a_x, i_y a_y, i_z a_z\}$, where  $(i_x,\;i_y,\;i_z) =  1, \cdots N$.
 As for the 2D case, we can perform  following unitary transformation, at each site of the lattice placing the unity $I=U^{-1}\times U$, with $U=\frac{1}{\sqrt{2}}\left({ }^1_1 { }^{-1}_{\;\;1}\right)$
 \bea R=U\otimes U\otimes U\otimes U\;\; W\;\;\; U^{-1}\otimes U^{-1}\otimes U^{-1}\otimes U^{-1}.\label{R}\ena
 The form of $R$ contrary to $W$ has the advantage, namely it contains only the elements for which the constraint $\sum_{i=1}^4\alpha_i=\sum_{i=1}^4\beta_i+{\rm{mod}}(2)$ does take place. We can present this $2^4\times 2^4$-dimensional matrix in this $2^2\times 2^2$ operator-matrix form:
 \bea
 R=\left(\ba{cccc}\textbf{R}_{00}^{00} &\textbf{R}_{00}^{01} &\textbf{R}_{00}^{10} &\textbf{R}_{00}^{11} \\\textbf{R}_{01}^{00} & \textbf{R}_{01}^{01}&\textbf{R}_{01}^{10} &\textbf{R}_{01}^{11} \\\textbf{R}_{10}^{00} &\textbf{R}_{10}^{01} &\textbf{R}_{10}^{10} &\textbf{R}_{10}^{11} \\\textbf{R}_{11}^{00} &\textbf{R}_{11}^{01} &\textbf{R}_{11}^{10} &\textbf{R}_{11}^{11}
 \ea\right)
 \ena
 where the operators $\textbf{R}_{i_1 i_2}^{j_1 j_2}$ themselves can be presented as $2^2\times 2^2$ matrices with corresponding statistical weights $ R_{{i_1}{i_2}{i_3}{
 i_4}}^{{j_1}{j_2}{j_3}{j_4}}$ by shifting the values of the indexes of $ R_{\bar{\sigma}_{\alpha_1}\bar{\sigma}_{\alpha_2}\bar{\sigma}_{\alpha_3}\bar{\sigma}_{
 \alpha_4}}^{\bar{\sigma}_{\beta_1}\bar{\sigma}_{\beta_2}\bar{\sigma}_{\beta_3}\bar{\sigma}_{\beta_4}}$  in (\ref{R}) as $i_k=(\bar{\sigma}_{{\alpha_k}}+1)/2$,  $j_k=(\bar{\sigma}_{{\beta_k}}+1)/2$.

 For the operators $\textbf{R}_{i_1 i_2}^{j_1 j_2}=\textbf{R}_{i_1 i_2}^{j_1 j_2}$ which have the property $\sum_{k=1}^2 i_k=\sum_{k=1}^2 j_k+{\rm{mod}}(2)$, the corresponding matrix also have the same structure, i.e.
 \bea
\textbf{R}_{i_1 i_2}^{j_1 j_2}=\left(\ba{cccc}{R}_{i_1 i_2
00}^{i_1 i_2 00} &0 &0 &{R}_{i_1 i_2 00}^{i_1 i_2 11} \\0 &
{R}_{i_1 i_2 01}^{i_1 i_2 01}&{R}_{i_1 i_2 01}^{i_1 i_2 10}
&0\\0&{R}_{i_1 i_2 10}^{i_1 i_2 01} &{R}_{i_1 i_2 10}^{i_1 i_2 10}
&0 \\{R}_{i_1 i_2 11}^{i_1 i_2 00} &0 &0&{R}_{i_1 i_2 11}^{i_1 i_2
11}
 \ea\right).
 \ena
 For the remaining matrix operators, when $\sum_{k=1}^2 i_k=\sum_{k=1}^2 j_k+1+{\rm{mod}}(2)$,
 correspondingly we can deduce
 \bea
 \textbf{R}_{i_1 i_2}^{j_1 j_2}=\left(\ba{cccc}0&{R}_{i_1 i_2 00}^{j_1 j_2 01} &{R}_{i_1 i_2 00}^{j_1 j_2 10} &0 \\{R}_{i_1 i_2 01}^{j_1 j_2 00}&0&0 &{R}_{i_1 i_2 01}^{j_1 j_2 11} \\{R}_{i_1 i_2 10}^{j_1 j_2 00}&0 &0 &{R}_{i_1 i_2 10}^{j_1 j_2 11} \\0 &{R}_{i_1 i_2 11}^{j_1 j_2 01} &{R}_{i_1 i_2 11}^{j_1 j_2 10} &0
 \ea\right).
 \ena
 This operator can be represented by means of the tensor products of the basic $2\times 2$ matrices, in terms of the generators of the algebra $sl(2)$, $\sigma^{z}=\{^1_0{}^{\;\;0\;}_{-1}\}$, $\sigma^{+}=\{^0_0{ }^1_0\}$, $\sigma^{-}=\{^0_1{}^0_0\}$ and the unity operator $I=\{^1_0{}^0_1\}$. Let us write the $R$-matrix in this operator form, where we have used following notations $\sigma_0^{0}=(I+\sigma^z)/2$, $\sigma_1^{1}=(I-\sigma^z)/2$, $\sigma_0^{1}=\sigma^+$, $\sigma_1^{0}=\sigma^{-}$ :
 \bea \label{R-s}
\mathbf{R}=R_{i_1 i_2 i_3 i_4}^{j_1 j_2 j_3
j_4}\sigma_{i_1}^{j_1}\otimes\sigma_{i_2}^{j_2}\otimes\sigma_{i_3}^{j_3}\otimes\sigma_{i_4}^{j_4},
 \ena
 and in the same time we can write also $\mathbf{R}=\sigma_{i_1}^{j_1}\otimes\sigma_{i_2}^{j_2}\textbf{R}_{i_1 i_2}^{j_1 j_2}$.

 \paragraph{The matrix elements of the operator $R$.}

The elements of the matrix $\textbf{R}_{i_1 i_2}^{j_1 j_2}$ are
presented explicitly below:

%
%
 We can note, that for this matrix there are the following
symmetry relations:
\bea R_{i_1 i_2 i_3 i_4}^{j_1 j_2 j_3 j_4}=R_{j_1 j_2 j_3
j_4}^{i_1 i_2 i_3 i_4}=R_{i_4 i_3 i_2 i_1}^{j_4 j_3 j_2
j_1}=R_{i_2 i_1 i_4 i_3}^{j_2 j_1 j_4 j_3}. \ena
And we can represent the following matrix elements by the following
expressions, where we have take $J_{1,2,3}=J_{y,x,z}$  and the
remaining elements can be found just from the above relations:
{\scriptsize \bea
&R_{0000}^{0000}=4(1-2\cosh{2J_x}\cosh{2J_y}\cosh{2J_z})+\cosh{4J_x}+\cosh{4J_y}+\cosh{4J_z}+
\cosh{4J_x}\cosh{4J_y}\cosh{4J_z},\nn&\\&
R_{0001}^{0001}=R_{0010}^{0010}=R_{0100}^{0100}=R_{1000}^{1000}=2\sinh{2J_z}(\cosh{2J_z}(1+\cosh{4J_x}\cosh{4J_y})-2\cosh{2J_x}\cosh{2J_y}),\nn
&\\&
R_{1000}^{0100}=4\sinh{2J_y}\sinh{2J_z}(\cosh{4J_x}\cosh{2J_y}\cosh{2J_z}-\cosh{2J_x}),\nn
&\\&
R_{0011}^{0000}=R_{0000}^{0011}=R_{0000}^{1100}=R_{1100}^{0000}=2\sinh{2J_y}(\cosh{2J_y}(1+\cosh{4J_x}\cosh{4J_z})-2\cosh{2J_x}\cosh{2J_z}),\nn
&\\&R_{0011}^{0011}=\cosh{4J_y}+\cosh{4J_z}-\cosh{4J_x}-2+
\cosh{4J_x}\cosh{4J_y}\cosh{4J_z},\nn&\\&
R_{0000}^{1010}=-4\sinh{2J_x}\cosh{2J_y}\cosh{2J_z}+(1+\cosh{4J_y}\cosh{4J_z})\sinh{4J_x},\nn&\\&
R_{0000}^{1001}=-4(\cosh{2J_z}-\cosh{2J_x}\cosh{2J_y}\cosh{4J_z})\sinh{2J_x}\sinh{2J_y},\nn&\\&
R_{0001}^{1000}=4\sinh{2J_x}\sinh{2J_y}\sinh{2J_z}(-1+2\cosh{2J_x}\cosh{2J_y}\cosh{2J_z}),\nn&\\&
R_{0010}^{1000}=4\sinh{2J_x}\sinh{2J_z}(-\cosh{2J_y}+2\cosh{2J_x}\cosh{4J_y}\cosh{2J_z}),\nn&\\
%
& R_{0101}^{0101}=-2+\cosh{4J_x}-\cosh{4J_y}+\cosh{4J_z}+
\cosh{4J_x}\cosh{4J_y}\cosh{4J_z},\nn&\\& R_{0100}^{0111}=
\cosh{4J_x}\sinh{4J_y}\sinh{4J_z},\nn&\\&
R_{0110}^{0110}=-\cosh{4J_x}-\cosh{4J_y}+\cosh{4J_z}+
\cosh{4J_x}\cosh{4J_y}\cosh{4J_z},\nn&\\&
R_{0110}^{0101}=(-1+\cosh{4J_x}\cosh{4J_z})\sinh{4J_y},\label{R-t}&\\&
R_{0111}^{0111}=4\cosh{2J_x}\cosh{2J_y}\sinh{2J_z}+(1+\cosh{4J_x}\cosh{4J_y})\sinh{4J_z},\nn&\\&
R_{1010}^{0101}=\cosh{4J_x}-\cosh{4J_y}-\cosh{4J_z}+
\cosh{4J_x}\cosh{4J_y}\cosh{4J_z},\nn&\\&R_{1001}^{0110}=2-\cosh{4J_x}-\cosh{4J_y}-\cosh{4J_z}-
\cosh{4J_x}\cosh{4J_y}\cosh{4J_z},\nn&\\&R_{1000}^{0111}=(-1+\cosh{4J_x}\cosh{4J_y})\sinh{4J_z},\nn&\\&
R_{1111}^{1111}=4(1+2\cosh{2J_x}\cosh{2J_y}\cosh{2J_z})+\cosh{4J_x}+\cosh{4J_y}+\cosh{4J_z}+
\cosh{4J_x}\cosh{4J_y}\cosh{4J_z},\nn&
\\&
R_{1011}^{0111}=4(\cosh{2J_x}+\cosh{4J_x}\cosh{2J_y}\cosh{2J_z})\sinh{2J_y}\sinh{2J_z},\nn&\\&
R_{0011}^{1111}=4\cosh{2J_x}\sinh{2J_y}\cosh{2J_z}+(1+\cosh{4J_x}\cosh{4J_z})\sinh{4J_y},\nn&\\&
R_{0000}^{1111}=-2+\cosh{4J_x}+\cosh{4J_y}-\cosh{4J_z}+\cosh{4J_x}\cosh{4J_y}\cosh{4J_z},\nn&\\&
R_{0011}^{1100}=-\cosh{4J_x}+\cosh{4J_y}-\cosh{4J_z}+\cosh{4J_x}\cosh{4J_y}\cosh{4J_z},\nn&\\&
R_{0011}^{1001}=\sinh{4J_x}(-1+\cosh{4J_y}\cosh{4J_z}),\nn&\\&
R_{0001}^{1011}=\sinh{4J_x}\cosh{4J_y}\sinh{4J_z},\nn &\\&
R_{0011}^{1010}=\sinh{4J_x}\sinh{4J_y}\cosh{4J_z},\nn&\\&
R_{0010}^{1011}=\sinh{4J_x}\sinh{4J_y}\sinh{4J_z},\nn&\\&
R_{1010}^{1111}=4\sinh{2J_x}\cosh{2J_y}\cosh{2J_z}+(1+\cosh{4J_y}\cosh{4J_z})\sinh{4J_x},\nn&\\&
R_{1011}^{1110}=4\sinh{2J_x}\sinh{2J_z}(\cosh{2J_y}+\cosh{2J_x}\cosh{4J_y}\cosh{2J_z}),\nn&\\&
R_{1011}^{1101}=4\sinh{2J_x}\sinh{2J_y}\sinh{2J_z}+\sinh{4J_x}\sinh{4J_y}\sinh{4J_z},\nn&\\&
R_{1001}^{1111}=4\sinh{2J_x}\sinh{2J_y}(\cosh{2J_z}+\cosh{2J_x}\cosh{2J_y}\cosh{4J_z}),\nn&
\ena}

 One can go to the Cardy's limit $2J_x\approx J_1 \Delta t$, $2J_y\approx J_2 \Delta t$, $e^{-2J_z}\approx h \Delta t$, with $\Delta t\ll 1$, in order to organize continuous limit in third direction, which can be regarded as time.  Thus, we can connect three dimensional statistical model with the quantum two dimensional problem, described by the corresponding Hamiltonian operator. As an example, the expansion of matrix element $R^{0000}_{0000}\;\;$ gives  $4(1-2\cosh{2J_x}\cosh{2J_y}\cosh{2J_z})+\cosh{4J_x}+\cosh{4J_y}+\cosh{4J_z}+
\cosh{4J_x}\cosh{4J_y}\cosh{4J_z}\approx 4(1-(h \Delta
t+\frac{1}{h \Delta t}))+2+((h \Delta t)^2+\frac{1}{h \Delta
t}^2)+ ((h \Delta t)^2+(\frac{1}{h \Delta t})^2)\approx
2(\frac{1}{h \Delta t})^2(1-2h\Delta t +O(\Delta t))$. In the leading order the
expansion of the $R-$matrix in its operator form is giving:
\bea R&=&2\frac{1}{(h \Delta t)^2}\left( I\otimes I \otimes I
\otimes I+\Delta t[J_1 (I\otimes \sigma_x\otimes I\otimes
\sigma_x+\sigma_x\otimes I\otimes\sigma_x\otimes
I)\right.\nn\\
&+&J_2(I\otimes I\otimes \sigma_x\otimes
\sigma_x+\sigma_x\otimes \sigma_x\otimes I\otimes I)\left.-h(I
\otimes I\otimes I\otimes \sigma_z\right.\nn\\
&+&\left. I\otimes
I\otimes\sigma_z\otimes I+I\otimes\sigma_z\otimes I \otimes
I+\sigma_z\otimes I\otimes I\otimes I)] \right). \ena
The operator in the parentheses coming with the coefficient
$\Delta t$ presents the cell Hamiltonian  for  2D
quantum spin model. Thus, the corresponding Hamiltonian defined on
square lattice reads
\bea
H&=&\sum_{i,j}\left(J_1[\sigma_x(2i,2j)\sigma_x(2i,2j+1)+\sigma_x(2i+1,2j)\sigma_x(2i+1,2j+1)]+\right.\nn\\
&+&\left. J_2[\sigma_x(2i,2j)\sigma_x(2i+1,2j)+\sigma_x(2i,2j+1)\sigma_x(2i+1,2j+1)]-h[\sigma_z(2i,2j)\nn\right.\\
&+&\left.\sigma_z(2i+1,2j)+\sigma_z(2i,2j+1)+\sigma_z(2i+1,2j+1)]\right)
\ena
At $J_1=0$ or $J_2=0$ this expression splits into the sum of two
quantum 1D Ising model's Hamiltonian operators defined on the
parallel chains (rows) of the square lattice.

\begin{figure}[t]
\unitlength=11pt
\begin{picture}(100,10)(5,-1)

\newsavebox{\rmatrix}

\sbox{\rmatrix}{\begin{picture}(10,0)(-2.5,5)
\put(30,4){\line(1,0){3}}\put(30,7){\line(1,0){3}}
\put(33,4){\line(0,1){3}}\put(30,4){\line(0,1){3}}
{\textcolor[rgb]{0.75,0.75,0.75}{\put(30,4){\line(2,1){2}}
 \put(32,5){\line(0,1){3}}\put(32,5){\line(1,0){3}}}}
\put(33,4){\line(2,1){2}}\put(35,5){\line(0,1){3}}
\put(30,7){\line(2,1){2}}\put(32,8){\line(1,0){3}}
\put(33,7){\line(2,1){2}}
\end{picture}}


%
\multiput(-22,0)(6,0){2}{\usebox{\rmatrix}}
\multiput(-18,8)(6,0){2}{\usebox{\rmatrix}}
\multiput(-23,4)(6,0){3}{\usebox{\rmatrix}}
\put(33,5){${R}_{1234}
\;=$}\put(5.5,5){\usebox{\rmatrix}}\put(38,3){\scriptsize$1$}
\put(41,3){\scriptsize$2$}\put(40,4.3){\scriptsize$3$}
\put(43,4.3){\scriptsize$4$} \put(38,7.3){\scriptsize$1'$}
\put(41,7.3){\scriptsize$2'$}\put(40,8.3){\scriptsize$3'$}
\put(43,8.3){\scriptsize$4'$}
\put(5.6,2.3){\scriptsize$\{2i,2j,2k\}$}
\put(5,6.8){\scriptsize$\{\!2i,\!2j,2k\!+\!1\!\}$}
\put(5,-2){\scriptsize$\{2i\!+\!1,2j\!-\!1,2k\!-\!1\}$}
\put(22,-2){(i)}\put(39,-2){(ii)}
\end{picture} \caption{$R$-matrix structure of 3D cubic lattice (i) and
 Cubic  ${R}_{1234}$-matrix (ii)}\label{fig1}
\end{figure}
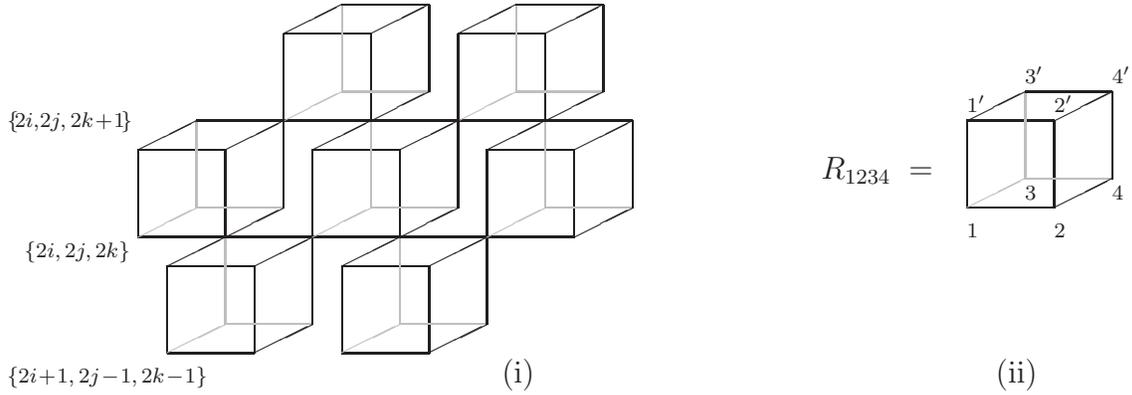

 \section{ $\mathcal{N}$-state chiral Potts model: the weight matrix, four-state vertex R-matrix nd 2D quantum Hamiltonian}
 \setcounter{equation}{0}

The $N$-state Potts model is the generalization of the IM, when at each site there are attached $\mathcal{N}$ spin variables:
$\bar{\sigma}=\{e^{\frac{i 2\pi \bar{n}}{\mathcal{N}}}\}$, $\bar{n}=0,1,...,\mathcal{N}-1$ mod $\mathcal{N}$. If to reformulate the product of the spins in the definition (\ref{wi}) by the following expression - $\bar{\sigma}_{\alpha}\bar{\sigma}_{\gamma}\Rightarrow \mathcal{N}(\delta(\bar{\sigma}_{\alpha}-\bar{\sigma}_{\gamma})-1/\mathcal{N})$ (which
is obviously an equality at $\mathcal{N}=2$), then we shall have the weight function of the $\mathcal{N}$-state ordinary Potts model defined
on the three dimensional cubic lattice
 \bea
 &&[{W^P}]_{\bar{\sigma}_{\alpha_1}\bar{\sigma}_{\alpha_2}\bar{\sigma}_{\alpha_3}\bar{\sigma}_{
 \alpha_4}}^{\bar{\sigma}_{\beta_1}\bar{\sigma}_{\beta_2}\bar{\sigma}_{\beta_3}\bar{\sigma}_{\beta_4}}=e^{H_1+H_2+H_3},\label{wp}\\
H_1&=&
 J_1 \mathcal{N}\left(\delta(\bar{\sigma}_{\alpha_1}-\bar{\sigma}_{\alpha_2})+\delta(\bar{\sigma}_{\alpha_3}-\bar{\sigma}_{\alpha_4})+
 \delta(\bar{\sigma}_{\beta_1}-\bar{\sigma}_{\beta_2})+\delta(\bar{\sigma}_{\beta_3}-\bar{\sigma}_{\beta_4})-\frac{4}{\mathcal{N}}\right),\nn
 \\
 H_2&=&J_2
 \mathcal{N}
 \left(
 \delta(\bar{\sigma}_{{\alpha_1}}
 -\bar{\sigma}_{\alpha_3})
 +\delta(\bar{\sigma}_{\alpha_2}-\bar{\sigma}_{\alpha_4})+
 \delta(\bar{\sigma}_{\beta_1}-\bar{\sigma}_{\beta_3})+\delta(\bar{\sigma}_{\beta_2}-\bar{\sigma}_{\beta4})-\frac{4}{\mathcal{N}}\right),\nn\\
 H_3&=&J_3 \mathcal{N}\sum_i^4\left(\delta({\bar{\sigma}}_{{\alpha_i}}-{\bar{\sigma}}_{{\beta_i}})-\frac{1}{\mathcal{N}}\right).\nn
 \ena

 The statistical sum is reproduced in the same way as in (\ref{Zi}).
 The two dimensional statistical model (if, e.g. $J_3$=0) at the self-dual point $(e^{J_1}-1)(e^{J_2}-1)=\mathcal{N}$ is the $Z_\mathcal{N}$ parafermionic Fateev-Zamolodchikov model which has
second order transition and can be described by conformal field theory (with $c=2(\mathcal{N}-1)/(\mathcal{N}+2)$) \cite{FZ}. An integrable generalization of this model is the 2D chiral Potts model, for which  2D vertex $R$-matrix has been constructed and which is satisfying the ordinary Yang-Baxter equations.

3D version of chiral Potts model can be constructed  in the following way. Note, that
\be\delta({\bar{\sigma}}_{{\alpha_1}}-{\bar{\sigma}}_{{\alpha_2}})=\delta(e^{\frac{2\pi i\bar{n}_{\alpha_1}k}{\mathcal{N}}}-e^{\frac{2\pi i\bar{n}_{\alpha_2}k}{\mathcal{N}}})\equiv \delta(\bar{n}_{\alpha_1}-\bar{n}_{\alpha_2}+{\textrm{mod}}{\mt N})=
\frac{1}{\mathcal{N}}\sum_{k=0}^{\mathcal{N}-1}e^{\frac{2\pi i k}{\mathcal{N}}(\bar{n}_{\alpha_1}-\bar{n}_{\alpha_2})}.\ee
 In chiral models there is assumed asymmetry, which allows to attach to each summand of this sum
with power $k$  it's own coupling coefficient $(J_a)^k$. E.g., for the 3D chiral Potts model the cell operators $H_r$ can be formulated as follows
\bea H_1&=&
 \sum_{k=1}^{\mathcal{N}-1}J_1^k\left(  e^{{\frac{2\pi i k}{\mathcal{N}}}(\bar{n}_{\alpha_1}-\bar{n}_{\alpha_2})}+e^{\frac{2\pi i k}{\mathcal{N}}(\bar{n}_{\alpha_3}-\bar{n}_{\alpha_4})}+
 e^{\frac{2\pi i k}{\mathcal{N}}(\bar{n}_{\beta_1}-\bar{n}_{\beta_2})}
 +
  e^{{\frac{2\pi i k}{\mathcal{N}}}(\bar{n}_{\beta_3}-\bar{n}_{\beta_4})}\right),\nn\\
   H_2&=&\sum_{k=1}^{\mathcal{N}-1}J_2^k\left(  e^{{\frac{2\pi i k}{\mathcal{N}}}(\bar{n}_{\alpha_1}-\bar{n}_{\alpha_3})}+e^{\frac{2\pi i k}{\mathcal{N}}(\bar{n}_{\alpha_2}-\bar{n}_{\alpha_4})}+
 e^{\frac{2\pi i k}{\mathcal{N}}(\bar{n}_{\beta_1}-\bar{n}_{\beta_3})}
 +
  e^{{\frac{2\pi i k}{\mathcal{N}}}(\bar{n}_{\beta_2}-\bar{n}_{\beta4})}\right),\nn\\
 H_3&=& \sum_{k=1}^{\mathcal{N}-1}\left(J_3^k \sum_{i=1}^4 e^{\frac{2\pi i k}{\mathcal{N}}(\bar{n}_{\alpha_i}-\bar{n}_{\beta_i})}\right).
 \ena

The vertex $R^P$-operator for  $\mathcal{N}$-state case can be obtained from the statistical weight in similarity with the Ising case (\ref{R}) by using the generalization of unitary $2\times 2$-operator $U$ to case of $\mt{N}\times \mt{N}$-operators $U^{\mt{N}}$,s which has the matrix elements

\be
{\left[U^{\mt{N}}\right]}_{k}^{p}=\frac{1}{\sqrt{\mt{N}}}e^{2\pi i \frac{(k-1)(p-1)}{\mt{N}}},\qquad k,\;p=1,...,\mt{N}.\ee

In order to reproduce the 2D quantum Hamiltonian corresponding to 3D chiral Potts model, we shall follow the logic of the works \cite{GSG}. We can involve $Z^{\mt N}$-symmetry operators $X,\; X^+$ and
$Z,\; Z^+$ . The operators act on the linear space with basis vectors $|k\rangle,\; k=0,\cdots \mt{N}-1$ as

\bea
Z_k^{p}=\delta_k^{p-1},\quad X_k^p= e^{\frac{2\pi i }{\mathcal{N}}}\delta_k^p, \quad
X^{\mt N}=1,\quad Z^{\mt{N}}=1,\qquad X Z=e^{\frac{2\pi i }{\mathcal{N}}}Z X.
\ena
Once we define the weights $W^x(n_a-n_b)$, $W^y(n_a-n_b)$ and $W^z(n_a-n_b)$ on the links which connect the vertexes $(a,b)$ with the state-variables
$\sigma_{a,b}=e^{\frac{2\pi i(n_{a,b})}{\mt{N}} }$ along the axes $x,\; y$ and $z$   as
\bea
W^x(n_a-n_b)=e^{\sum_{k=1}^{\mt{N}-1} {J_1^k e^{\frac{2\pi i}{\mt{N}}(n_a-n_b) k}}},\nn\\
W^y(n_a-n_b)=e^{\sum_{k=1}^{\mt{N}-1} {J_2^k e^{\frac{2\pi i}{\mt{N}}(n_a-n_b) k}}},\\
W^x(n_a-n_b)=e^{\sum_{k=1}^{\mt{N}-1} {J_3^k e^{\frac{2\pi i}{\mt{N}}(n_a-n_b) k}}},\nn
\ena
we can reformulate the weight matrices in a following way
\bea
&&[W^P]_{\bar{n}_{\alpha_1}\bar{n}_{\alpha_2}\bar{n}_{\alpha_3}\bar{n}_{
 \alpha_4}}^{\bar{n}_{\beta_1}\bar{n}_{\beta_2}\bar{n}_{\beta_3}\bar{n}_{\beta_4}}=\\\nn
&& W^x(\bar{n}_{\alpha_1}-\bar{n}_{\alpha_2})W^x(\bar{n}_{\alpha_3}-\bar{n}_{\alpha_4})
W^x(\bar{n}_{\beta_1}-\bar{n}_{\beta_2})W^x(\bar{n}_{\beta_3}-\bar{n}_{\beta_4})\times\\\nn
&&W^y(\bar{n}_{\alpha_1}-\bar{n}_{\alpha_3})W^y(\bar{n}_{\alpha_2}-\bar{n}_{\alpha_4})
W^y(\bar{n}_{\beta_1}-\bar{n}_{\beta_3})W^y(\bar{n}_{\beta_2}-\bar{n}_{\beta_4})\times\\\nn
&&W^z(\bar{n}_{\alpha_1}-\bar{n}_{\beta_1})W^z(\bar{n}_{\alpha_2}-\bar{n}_{\beta_2})
W^z(\bar{n}_{\alpha_3}-\bar{n}_{\beta_3})W^z(\bar{n}_{\alpha_4}-\bar{n}_{\beta_4}).
\ena
For the reformulation of the $R^P$-operator in terms of $Z,\; X$ matrices, let us define the following link transfer operators:  horizontal $ S^{1,2}$ transfer operator along the axes $x,y$ and the vertical transfer operator  $T$ along the axe $z$.
\bea
&&S^{1,2}=\sum {\bar{W}}^{x,y}(k)(X\otimes X^+)^k,\quad \bar{W}^{x,y}(k)=\frac{1}{\mt{N}}W^{x,y}(n)e^{\frac{2\pi i}{\mt{N}}(n k)},\\
&&W^{x,y}({\bar{n}}'-{\bar{n}})=\langle \bar{n}|\otimes{\bar{n}}'|S^{1,2}|{\bar{n}}'\rangle\otimes |\bar{n}\rangle,\\
&&T=\sum W^z(k) Z^k,\quad W^z({\bar{n}}'-{\bar{n}})=\langle \bar{n}|T |\bar{n}'\rangle.
\ena
Then
\bea
[R^P]_{1234}=\left[[S^1]_{12} ([S^2]_{13}\otimes [S^2]_{24}) [S^1]_{34}\right]\left(T_{1}\otimes T_2\otimes T_3\otimes T_4 \right)
\left[[S^1]_{12} ([S^2]_{13}\otimes [S^2]_{24}) [S^1]_{34}\right].\nn\\
\ena
After implementation of the unitary transformations by means of the mentioned operators, which is actually  Fourier transformation of the vector basis, the matrix forms of the operators $X$ and $Z$ are interchanging their view, and now the
operator $Z$ is diagonal.

 As in the case of 3DIM we can take the Cardy's limit  at $\Delta t\ll 1$ for this generalized case and construct the 2D quantum lattice Hamiltonian for the chiral Potts model.
 %
%
\be R^P=I\otimes I \otimes I \otimes I+\Delta t H^P.\ee
The Hamiltonian will have  similar to the 1D quantum chain chiral Potts Hamiltonian view \cite{GSG}

\bea
H^P&=&\sum_{i,j}\sum_{k=1}^{\mt{N}-1}\left({J}_{x k}\left[X^k(2i,2j){X}^{+k}(2i,2j+1)+
X^k(2i+1,2j){X^+}^k(2i+1,2j+1)\right]\right.\nn\\
&+&\left. J_{y k}\left[X^k(2i,2j)X^{+k}(2i+1,2j)+X^k(2i,2j+1)X^{+k}(2i+1,2j+1)\right]-h_k[Z^k(2i,2j)\nn\right.\\
&+&\left.Z^k(2i+1,2j)+Z^k(2i,2j+1)+Z^k(2i+1,2j+1)]\right)\nn\\
\ena
The Hamiltonian of this basis has the symmetry $[H^p,\mathbb{Z}^P]=0$ with charge $\mathbb{Z}^P=\prod_{i,j}^{N,N} Z{(i,j)}$.

 \section{$\mathcal{N}=2$: The fermionic representation: Free fermionic conditions}
\setcounter{equation}{0}

 In the article \cite{SA} we have represented the 2d model in terms of the graded
 basis \cite{GM}, associating with each site of the lattice a pair of the creation and annihilation fermionic operators (so called "0"-spin fermions),
  $c^+_{\alpha},\;c_{\alpha}$,
   $c^+_{\alpha}c_{\alpha'}+c_{\alpha'}c^+_{\alpha}=\delta_{\alpha\alpha'}$.
    Then the operators defined above $\sigma^i_j$ in the  Fock space with basis
    $|0\rangle,|1\rangle=c^+|0\rangle$ can be represented in terms
     of the fermionic operators,
  \bea
  \sigma^0_1=|0\rangle\langle 1|=c,\quad \sigma^1_0=|1\rangle\langle 0|=c^+,\quad\sigma^0_0=|0\rangle\langle 0|=1-c^+c,\quad\sigma^1_1=|1\rangle\langle 1|=c^+c.
  \ena
 This is the reflection of the spin-fermion correspondence (Jordan-Wigner transformation) \cite{JW}.  In the lattices with definite arrangement of the sites at
 which the spin operators $\sigma^{\pm,z}_{\alpha}$ are attached,
  the Jordan-Wigner transformation is non-local, in order to ensure
  the anti-commutation behavior of the fermionic operators at different
   sites:
 \bea
 \sigma^+_{\alpha}=\prod_{\gamma=1}^{\alpha-1}\left[1-2 c^+_{\gamma}c_{\gamma}\right]c^+_{\alpha},\quad
 \sigma^-_{\alpha}=\prod_{\gamma=1}^{\alpha-1}\left[1-2 c^+_{\gamma}c_{\gamma}\right]c_{\alpha},\quad
 \sigma^z_{\alpha}=2 c^+_{\alpha}c_{\alpha}-1.
 \ena
For the three dimensional cubic lattice the variable $\alpha$
denotes the vertices labelled with the integers $\{i,\;j,\;k\}$
corresponding to the coordinates $\{x,\; y,\;z\}=\{ia,\; ja,\;
ka\}$ - where $a$ is the lattice spacing.

The  3D R-matrix (\ref{R-s}) under consideration can be expressed by fermionic operators in accordance with approach
developed in articles \cite{Avakyan-1997, GM, Sedrakyan-2000, Sedrakyan-2001}
and  adapted for evaluating the partition functions (2D IM, XY cases) in \cite{SA}. As a result we shall have
the following graded formulae for the operator (\ref{R-s}) defined on the
space
 $V_1\otimes V_2 \otimes V_3 \oplus V_4$, $V_k = \{0\rangle_k \; |1\rangle_k\}$,
\bea
  R_{1234} = \sum_{
i_k=1,j_k=1[k=1,2,3,4]} R^{j_1 j_2 j_3 j_4}_{i_1 i_2 i_3 i_4}
|j_1\rangle_1 |j_2\rangle_2 |j_3\rangle_3 |j_4\rangle_4
 { }_4\langle i_4| { }_3\langle i_3| { }_2\langle i_2| { }_1\langle i_1|= \label{R-p}\\\nn
 = \sum_{
i_k=1,j_k=1[k=1,2,3,4]} R^{j_1 j_2 j_3 j_4}_{i_1 i_2 i_3
i_4}(-1)^{p(R)} |j_1\rangle{ }_1\langle i_1| |j_2\rangle{ }_2\langle
i_2| |j_3\rangle{ }_3\langle i_3| |j_4\rangle{ }_4\langle i_4|
\ena
 Here the factor $(-1)^{p(R)}$
indicates, that in fermionic representation the
Fock space is graded, and the states $|0 \rangle_i,\;|1\rangle_j$
have different gradings: the states $|0 \rangle_i$ with different
$i$ are commutative with one another and with the states
$|1\rangle_j$, and they have the parity $p(0) = 0$, meanwhile the
states $|1\rangle_i$ with different $i$ are anti-commute and have
the parity $p(1) = 1$. This means $|a_i\rangle_i |a_j\rangle_j
=|a_j\rangle_j |a_i\rangle_i(-1)^{p(a_i)p(a_j)}$. Thus we can check
the parity $p(R)$ of the $R$-operator in the relation
(\ref{R-p}),
\bea p(R)=\sum_{t=1}^3 p(i_t)\sum_{k=t+1}^4(p(i_k)+p(j_k)). \ena

The fermionic representation of the discussed $R$-matrix is a local
operator, as it is even operator in terms of the fermionic
operators, and it means that the non-local term of the Jordan-Wigner
transformation must be counted even times,  and thus must be
reduced, as $(1-2n)(1-2n)=1$. What is the advantage of the fermionic
representation - it gives an opportunity to represent the
statistical sum (partition function) and the other statistical quantities as integrals with respect to the fermionic variables.
This can be achieved by means of the
coherent basis of the fermionic operators formulated via the
Grassmann variables $\psi_i,\;\bar{\psi_i}$ (\ref{psi}), which  fulfill orthonormality and
 completeness relations, see Appendix {A}(\ref{psi-oc}).

So we can represent the R-matrices in the form of
$R=A_0:e^{A(\bar{c}c)}:$, and in the general case the fermionic
action for the elementary cell of the cubic lattice can have
interaction terms up to the  8th degree $A=\sum_{i=1}^4 A_i[c'
c]^{i} $, where $c',\; c$ are the fermionic operators from the set
$c_i,\; \bar{c}_j$ situated on the sites of the cube. It happens, that for the
case of 2D Ising model the fermionic action
contains only quadratic terms  \cite{SA} and so describes free
fermions \cite{Baxt78}.

 Then one can represent the partition function of the model
defined on the 3D cubic lattice as integral representation over the
 fermionic lattice action
\bea Z=\prod_{3D}R=(A_0)^{N^2}
 \int D\psi D\bar{\psi} e^{\sum_{3D}\mathbf{ A}(\bar(\psi),\psi)-\sum
 \bar{\psi}\psi}.
\ena

 This
can be achieved  by writing in the partition function all the
R-matrices in terms of the coherent basis, situating the
unity operators in the operator form (\ref{i-p}) at  each vertex of
the 3D cubic lattice, and then taking the trace (\ref{int}).

In the coherent basis the cell action for the cube R-matrix which
acts on the vector spaces on the square (see the figure \ref{fig1})
with the vertices noted by $\{1,2,3,4\}$, has the following form
\be
\langle \bar{\psi}_4| \langle \bar{\psi}_3|
 \langle \bar{\psi}_2| \langle \bar{\psi}_1|R|\psi_1\rangle|\psi_2\rangle|\psi_3\rangle
 |\psi_4\rangle=A_0 e^{A(\bar{\psi},\psi)+\sum_i
\bar{\psi}_i\psi_i},\ee
 where
\bea
\label{A}
A &=& A_2 + A_4 + A_6 + A_8, \\
 A_2 &=& \sum_{i,j=1}^4 a^i_j \bar{\psi}_i\psi_j+ \sum_{i<j}^4 a_{ij} {\psi}_i\psi_j+ \sum_{i<j=1}^4 a^{ij}\bar{\psi}_i\bar{\psi}_j,\\
 A_4&=&\sum_{i,j,k,r=1}^4\left(a^{ij}_{kr}\bar{\psi}_i\bar{\psi}_j
 {\psi}_k\psi_r+a^{ijk}_{r}\bar{\psi}_i\bar{\psi}_j
 \bar{\psi}_k\psi_r+a^{i}_{jkr}\bar{\psi}_i{\psi}_j
 {\psi}_k\psi_r\right)+\nn\\
 &+& a^{1234}\bar{\psi}_1\bar{\psi}_2
 \bar{\psi}_3{\bar{\psi}}_4+a_{1234}{\psi}_1\psi_2
 {\psi}_3\psi_4,\\
 A_6&=&\sum_{i,j,k,r,p,t=1}^4\left(a^{ij}_{1234}\bar{\psi}_i\bar{\psi}_j
 {\psi}_1\psi_2
 {\psi}_3\psi_4+a^{1234}_{pt}\bar{\psi}_1\bar{\psi}_2
 \bar{\psi}_3{\bar{\psi}}_4\psi_p\psi_t+a^{ijk}_{rpt}\bar{\psi}_i\bar{\psi}_j
\bar{\psi}_k\psi_r\psi_p\psi_t\right),\\
A_8&=&a_{1234}^{1234}\bar{\psi}_1\bar{\psi}_2
 \bar{\psi}_3{\bar{\psi}}_4 {\psi}_1\psi_2
 {\psi}_3\psi_4.
\ena    %

 Comparing the expressions of two realizations of the R-matrix we can easily find the relations between the coefficients $a_{---}^{---}$ and the matrix elements $R_{---}^{---}$.
Particularly:
\bea
A_0 &=& R_{0000}^{0000},\;
\qquad a_{ij}
= R^{0000}_{-i-j-}(-1)^p/A_0, \qquad  a^{ij} = R_{0000}^{-i-j-}(-1)^p/A_0,\nn\\ \nn\\
 a_i^j &=& R_{-i-}^{-j-}(-1)^p/A_0-\delta_i^j,\qquad
 a_{ij}^{kr}=R_{-i-j-}^{-k-r-}(-1)^p/A_0+\delta_i^k
 \delta_j^r+R_{-i-}^{-k-}\delta_j^r(-1)^p/A_0+ \nn\\
 &+&R_{-i-}^{-r-}\delta_j^k(-1)^p/A_0
 + R_{-j-}^{-k-}\delta_i^r(-1)^p/A_0
 -(a_j^r a_i^k(-1)^p+a_{ij}a^{kr}(-1)^p-
 a_i^r a_j^k(-1)^p),\nn \\
 a_{1234}&=&R_{1111}^{0000}/A_0-a_{12}a_{34}+a_{13}a_{24}-a_{14}a_{23},\nn\\
 a^{1234}&=&R^{1111}_{0000}/A_0-a^{12}a^{34}+a^{13}a^{24}-a^{14}a^{23},\\
 a_{ijk}^r&=&R_{-i-j-k}^{-r-}(-1)^p/A_0-(a_{ik} a_j^r(-1)^p+a_{ij}a_k^{r}(-1)^p-
 a_{jk} a_i^r(-1)^p),\nn\\
  a^{ijk}_r&=&R^{-i-j-k}_{-r-}(-1)^p/A_0-(a^{ik} a^j_r(-1)^p+a^{ij}a^k_{r}(-1)^p-
 a^{jk} a^i_r(-1)^p),\nn\\
  a_{ijk}^r&=&R_{-i-j-k}^{-r-}(-1)^p/A_0-(a_{ik} a_j^r(-1)^p+a_{ij}a_k^{r}(-1)^p-
 a_{jk} a_i^r(-1)^p)\nn
\ena
The expressions for the elements $A_6$ and $A_8$ can be easily deduced in the same manner.
 Here by $R_{-i-}^{-j-},\; R^{0000}_{-i-j-},\; R_{0000}^{-i-j-},...
$ we denote the matrix elements, for which all the indexes are
$0$, besides of those, which are at the positions $i, j, ...$,
e.g. $R_{-1-}^{-4-}=R_{1000}^{0001}$. The sign $(-1)^p$ is 
taking into account the grading. The parity of each summand in
the expressions must be checked separately. Here we can write
explicitly some of the $a$-coefficients. At first, the
coefficients of $A_2$ easily can be derived from (\ref{R-t}). It is clear that the $a_i^j$ have the same symmetries, as the matrix
elements $R_{-i-}^{-j-}$ in (\ref{R-t}), e.g.
$a_1^1=a_2^2=a_3^3=a_4^4$, $a_1^2=a_3^4$, and so on. The
interaction terms $A_k,\; k>2$ also can be exactly calculated. Particularly,
some of the expressions in $A_4$ are identically null, such as the terms $a_{12}^{12}=0$, $a_{34}^{34}=0$, $a_{13}^{13}=0$,
$a_{24}^{24}=0$, $a_{1234}=a^{1234}=0$, meanwhile  the
expressions of the following terms are
\bea
a_{23}^{23}&=&-a_{14}^{14}=-a_{14}^{23}=16(\sinh{2J_1}\sinh{2J_2}\sinh{2J_3})^2/A_0^2\\
a_{1}^{234}&=&-16(\sinh{2J_1}^2\sinh{2J_2}^2\sinh{2J_3})(\cosh{2
J_1\cosh{2J_2}}-\cosh{2J_3})/A_0^2\ena

 From the checking all the expressions in the sets $A_4$ ($A_6,\; A_8$) it follows
that free fermionic condition (i.e. $A_{k>2}=0$) means
\be \sinh{J_1}\sinh{J_2}\sinh{J_3}=0, \label{freeR}\ee
 as all the functions in
$A_4$ are proportional to $\sinh{J_1}\sinh{J_2}\sinh{J_3}$. This is the case of 2DIM. The statistical sum of any free model can be evaluated simply by direct calculations, particularly using the  Fourier transformation in the Grassmann variable's space.

For being precise, before performing the fermionic transformation of the $R$-matrix, one should define at first the non-check graded $\bf{R}$ matrix, as
\be {\bf{R}}_{i_1 i_2 i_3 i_4}^{j_1 j_2 j_3
j_4}=(-1)^{\sum_{k=4}^2 (p_{j_k}\sum_{t=k-1}^1 p_{j_t})} R_{i_1
i_2 i_3 i_4}^{j_4 j_3 j_2 j_1}, \ee
which means, that in the formulas for $A_k$ we must take into account the
following transformations $j_i\to j_{5-i}$ for the upper indexes, and the corresponding
changes in the signs.

\paragraph{Critical point of the model in the free fermionic limit, for
small coupling constants}

 As it is known, the critical point of the
three-dimensional Ising model is described by a conformal field
theory \cite{Pol}, and the conformal field theory is under active
investigation using the method of the conformal bootstrap
\cite{Beta,IMc}. By means of this method and  by Monte Carlo
simulations there are obtained rather precise information about
the critical exponents. For the homogeneous 3D IM the best known
critical value the of coupling constant is $0.22165455$.

  We also can try to find the critical points in the free fermionic limit. This means, that we must take only the quadratic part of the action in (\ref{A}), which is justified at small $J_i$-s, as we can see from the exact values of the coefficients in the terms $A_4,\;A_6$ and $A_8$. Then, following to the steps in \cite{SA}, where an exact calculations has been done for 2DIM, we can perform a Fourier transformation of the fermionic basis in 3D lattice with antiperiodic boundary conditions. After redefining the Grassmann fields at the half of the momenta space as $\bar{\psi}_i(\pi-p_x,\pi-p_y,\pi-p_z)=\psi_{i+4}(p_x,p_y,p_z)$  and
   ${\psi}_i(\pi-p_x,\pi-p_y,\pi-p_z)=-\bar{\psi}_{i+4}(p_x,p_y,p_z)$ ($i=1,2,3,4$), we can represent the partition function as a product of the determinants of $8\times 8$ matrices.
   The zeroes of the partition function
 give the approximated value of the coupling constant, which is $J_c^f=0.270325$ in the free fermionic limit.

\section*{Summary and Acknowledgements}

 In this work we have presented the 3D generalizations of the 2D integrable models - IM and $\mt{N}$-state chiral Potts model in the vertex four-state $R$-matrix formulation. This will give an advantage in the theoretical (in the framework of 3D ABA) and numerical investigations of these models or their modifications.

 The work was supported by the Science Committee of RA, in the frames of research projects  20TTWS-1C035 and  20TTAT-QTa009.


\section{Appendix}
\setcounter{equation}{0}
\paragraph{A: Coherent Basis and Grassmann variables}
\bea \label{psi} c_i|\psi_i\rangle=\psi_i|\psi_i\rangle,&\quad &
\langle \bar{\psi}_i|c^+_i=\langle \bar{\psi}_i|\bar{\psi}_i,\\
\label{psi-oc}
\langle\bar{\psi}_i|\psi_j\rangle=\delta_{ij}e^{\bar{\psi}_i\psi_i}\quad,&\quad&
\int d\bar{\psi}_id\psi_i
e^{-\bar{\psi}_i\psi_i}|\psi_i\rangle\langle \bar{\psi}_i| =
I.\;\;\; \ena

%
%

Any operator $K(\{c^+_i,c_j\})$ in the fermionic coherent basis reads as
\bea \mathcal{K}(\{\bar{\psi}_i,\psi_j\})\equiv\langle \prod
\bar{\psi}_i|K(\{c^+_i,c_j\})|\prod \psi_j\rangle=e^{\sum_i
\bar{\psi}_i\psi_i} K(\{\bar{\psi}_i,\psi_j\}). \ena
The trace of the operator $K(\{c^+_i,c_j\})$ in coherent-states is
an integral over the Grassmann variables,

\bea \mathrm{tr} K(\{c^+_i,c_j\})= \int D\psi D\bar{\psi}
e^{\sum_i\bar{\psi}_i\psi_i}
\mathcal{K}(\{\bar{\psi}_i,\psi_j\}),\quad \quad D\psi
D\bar{\psi}=\prod_i d\psi_i d\bar{\psi}_i.\label{int}
 \ena
The following integral representation takes place for the identity
operator.
\be I=\int d\bar{\psi}(i,j)d\psi(i,j)
e^{-\bar{\psi}(i,j)\psi(i,j)}|\psi(i,j)\rangle\langle
\bar{\psi}(i,j)|.\label{i-p}\ee

\paragraph{B: Local integrability equations for R-matrices defined on the cube}

One can use the following  simplest extension of the Baxter's
transformations \cite{Baxt} for proper parametrization of the
$R(u,w)$-matrix elements in order to check the integrability
properties in the context of the Bethe Ansatz.
\bea &e^{\pm 2J_1}=\mathrm{cn}[i \;u,k]\mp i
\;\mathrm{sn}[i\;u,k],&\nn
 \\&e^{\pm 2J_2}=\mathrm{cn}[i \;w,k]\mp i \;\mathrm{sn}[i\;w,k],&\nn \\
&e^{\pm 2J_3}=i (\mathrm{dn}[i \;(u+w),k]\pm 1)/(k
\;\mathrm{sn}[i\;(u+w),k]).&\label{tranB}
\ena
When $J_1=0$or $J_2=0$ these relations are equivalent to the
corresponding formulas for the 2d case. Another possible variation
of the Baxters
transformation could have such kind expression for
$J_3$:
\bea
 e^{\pm 2J3} =i (\mathrm{dn}[i \;(u),k]+\mathrm{dn}[i \;(w),k]\pm 1)/(k
\;\mathrm{sn}[i\;(u+w),k])\ena

By these transformations for 2d case one ensures the form of the
$R(u)$ matrix satisfying Yang-Baxter equations (YBE) with additive
spectral parameter -
 $$R(u_{12})R(u_{13})R(u_{23}) = R(u_{23})R(u_{13})R(u_{12}),$$
where $u_{ij}=u_i-u_j$.

The symbolic extension of this relation for 3D case with $R(u, v, u
+ v)$-matrix has the form
\bea
 R(u_{12}, u_{51}, u_{52})R(u_{34}, u_{53}, u_{54})R(u_{36},
u_{13}, u_{16})R(u_{46}, u_{24}, u_{26}) = \\\nn R(u_{46},u_{24}, u_{26})R(u_{36}, u_{13},
u_{16})R(u_{34}, u_{53}, u_{54})R(u_{12} u_{51}, u_{52}).
\ena

 The spectral parameter
dependence here is taken as for the standard 3D vertex R-matrix,
defined on the tensor product of three vector spaces $V_i\otimes
V_j \otimes V_k$, for which the local integrability conditions are
the vertex version of ZTE or Semi-Tetrahedral equations (see
\cite{ZZ}, \cite{AShKhS}). The spectral parameters are attached to the three  lines orthogonal to the faces of the cubes.
   If to check the cube equations taking the constructed $R$-matrices, of
course, there are solutions to these equations,  which correspond to the situations equivalent
to the 2D case - $J_{1,2}=0$, when the cube equations transform to the set
of YB equations. However for general case one can take the
suggested parametrization as a starting point, and look for the
solutions after modifications both of the R-matrices and
local equations.

Note, that for the cube equations also one can suggest restricted
variant of the equations - simplified cube equations, where as
intertwiner matrtices one can take two-particle $R_{ij}$ matrices,
as in \cite{AShKhS}.

\paragraph{C: Integrable 3D model with general R matrix of non-symmetric
free-fermionic structure: Commutativity of the transfer matrices.}

One can note, that if the given R-matrix of any dimensional
statistical model has a such structure that it lets possible to represent the statistical sum
as a generating functional with free particle action, then the model is integrable.  Such
model is an integrable model, but however it does not mean that in
Bethe anzats framework such R-matrix satisfies to a local
integrability condition, or transfer matrices with different spectral parameters 
commute
and there is an intertwiner matrix ensuring it.
 For the known free-fermionic cases ($XX,\; XY$ or 2D IM)
 the R-matrix itself has a similar structure $R_f \equiv : e^{Ac'c} :$. For example, for the most and entirely investigated 2D case, the
most general form $R_f =:e^{\sum^2_{i,j=1} a^j_i \bar{c}_i c_j
+\sum^2_{i>j} a^{ij} \bar{c}_i \bar{c}_j+\sum^2_{i>j} a_{ij}{c}_i c_j}:$
 has arbitrary coefficients, meanwhile YBE
solutions put definite restrictions on them, see e.g \cite{Baxt,SA} for
homogeneous YBE, and \cite{SKS,SHKH} for inhomogeneous YBE.  The periodic quadratic  operators can be easily  diagonalized in the Fourier  transformation basis.

The
free-fermionic conditions for 3D matrix with standard vertex
structure $R_{ijk}$ (the vector states are situated on the six
links) is presented in \cite{AShKhS}, and a solution to semi-tetrahedral
equations is presented therein. The free fermionic 3D models are considered also in \cite{BS}.  As we  have seen in this article the free-fermionic condition of
the 3D IM brings to the relation (\ref{freeR}). In general case the free-fermionic conditions for
the cube $R_{ijkr}$-matrix (the vector states are on the eight
vertices) can be defined in similar manner, expanding the
corresponding exponent in the normal ordered form and comparing
the matrix elements. As example we can present a relation
\bea R_{0000}^{0000}R_{1010}^{1010} =
R_{1000}^{1000}R_{0010}^{0010} +
 R_{1010}
^{0000}R_{0000}^{1010}-R_{0010}^{1000}R_{1000}^{0010}.\ena

 In fact, such kind equations, as in \cite{AShKhS},  mean
the equalities between the appropriate matrix-minors in the R-matrix.

\end{document}